\def\gtorder{\mathrel{\raise.3ex\hbox{$>$}\mkern-14mu
\lower0.6ex\hbox{$\sim$}}}
\def\ltorder{\mathrel{\raise.3ex\hbox{$<$}\mkern-14mu
\lower0.6ex\hbox{$\sim$}}}
\documentstyle[epsf,rotate]{l-aa}
\begin{document}
\thesaurus{03(11.16.1; 11.19.2;
11.09.4; 09.04.1)}

\title{Are spiral galaxies optically thin or thick?}

\author{E. M. Xilouris \inst{1,2}
\and Y. I. Byun \inst{3}
\and N. D. Kylafis \inst{4,2}
\and E. V. Paleologou \inst{4,2}
\and J. Papamastorakis \inst{4,2}}

\offprints{xilouris@physics.uch.gr}

\institute{
University of Athens, Department of Physics, Section of
Astrophysics, Astronomy \& Mechanics, 157 83 Athens, Greece
\and Foundation for Research and Technology-Hellas, P.O. Box 1527,
711 10 Heraklion, Crete, Greece
\and Center for Space Astrophysics \& Department of Astronomy,  
Yonsei University, Seoul 120-749, Korea
\and University of Crete, Physics Department, P.O. Box 2208,
710 03 Heraklion, Crete, Greece}

\date{Received date / Accepted date}
\maketitle
\markboth{E. M. Xilouris et al.: Are spiral galaxies optically thin or
thick?}{}
\begin{abstract}
The opacity of spiral galaxies is examined by modelling the dust
and stellar content of individual galaxies. The three dimensional
model that we use assumes exponential distributions for
the dust and the stars in the disk, while the $R^{1/4}$ law
is used to describe the bulge. In this model, both
absorption and scattering by the dust are taken into account.
The model is applied to five late-type spiral galaxies
(NGC 4013, IC 2531, UGC 1082, NGC 5529 and NGC 5907) using their
optical (and near infrared for IC 2531) surface photometry.
For these galaxies we have determined the scalelengths and scaleheights of the
stars and the dust in the disk, the bulge characteristics,
the inclination angle and the face-on optical depth.
Computation of the dust masses, as well as the extinction
as a function of the wavelength, are also reported.

Having analyzed a total of seven galaxies
thus far, the five galaxies mentioned
above plus UGC 2048 and NGC 891 presented in 
(Xilouris et al. 1997, 1998),
we are able to draw some general
conclusions, the most significant of which are:\\
1) The face-on central optical depth is less than one in all optical bands,
indicating that typical spiral galaxies like the ones
that we have modelled would be completely transparent if they were to be seen face-on.\\
2) The dust scaleheight is about half that of the stars, which means
that the dust is more concentrated near the plane of the disk.\\
3) The dust scalelength is about 1.4 times larger than that of the stars 
and the dust is more radially extended than the stars.\\
4) The dust mass is found to be about an order of a magnitude
more than previously measured using the IRAS fluxes, indicating the existence
of a cold dust component.
The gas-to-dust mass ratio calculated
is close to the value derived for our Galaxy.\\
5) The derived extinction law matches quite well the Galactic extinction
law, indicating a universal dust behaviour.
\end{abstract}


\section{Introduction}
The amount and the way that dust is distributed within spiral galaxies
as well as its extinction effects on the galactic starlight
has been a matter of debate over the last few years. Since Disney,
Davies and Phillipps (1989) proposed that spiral galaxies may be optically thick,
a long series of papers has been published by different authors
(see e.g. the introduction of Xilouris et al. 1997) with
conclusions that often differ widely.
Recent papers tend to support the idea of optically thin/moderately
opaque spiral galaxies. This means that galaxies are either
optically thin throughout their disks or that they are optically thin
in the outer regions and moderately opaque at the center (e.g.
Giovanelli et al. 1994, 1995;
Kodaira \& Yamashita 1996; R\"{o}nnback \& Shaver 1997; Domingue et al.
1998; Moriondo et al. 1998; Kuchinski et al. 1998; Gonz\'{a}lez
et al. 1998).
On the other hand, recent far infrared/submillimeter observations
of spiral galaxies give a direct picture of the dust distribution
and a good estimate of the total amount of dust (e.g. Alton et al. 1998a, 1998b;
Haas et al. 1998; Bianchi et al. 1998; Kr\"{u}gel et al. 1998).
These observations indicate that the cold dust component ($\sim 10-20$ K) that is now 
detected, boosts the dust masses about an order of magnitude higher than
previously estimated by IRAS observations. On the other hand, a more radially
extended dust distribution compared to that of the stars is derived.

In our studies
we perform detailed modelling of spiral galaxies, aiming at the
determination of the parameters that describe the three-dimensional dust
distribution in these galaxies.
For the three dimensional distribution of stars and dust
we use simple and smooth geometries (e.g. axisymmetric exponentials
for the stars and the dust in the disk and the $R^{1/4}$ law
for the bulge). There is no doubt that in reality spiral galaxies
are clumpy with small and large scale structure
(e.g. spiral arms with the dust trailing the arms in a patchy
way). One has to keep in mind those oversimplifications. The edge-on
picture though that we use cancels out most of the variation along the
line of sight and allows us to use simple mathematical
functions in order to get a mean description of the galaxies. 
We have verified this with detailed computations
(Misiriotis et al. in preparation) which take into account 
the spiral structure.

The two galaxies (UGC 2048 and NGC 891) that we have already
modelled (Xilouris et al. 1997; hereafter Paper I and 
Xilouris et al. 1998; hereafter Paper II)
using the radiative transfer
model of Kylafis \& Bahcall (1987; hereafter KB87),
show low values of the
optical depth in the face-on orientation, indicating low
opacities for these galaxies, while the predicted dust masses
and the dust distributions are now supported by the far infrared/submillimeter
observations mentioned above. Our aim is to continue this kind
of study with as many galaxies as possible in various orientations
and thus improve the statistics. 
In this paper we do the analysis of the edge-on late-type spiral
galaxies NGC 4013, IC 2531, UGC 1082, NGC 5529 and
NGC 5907 (see Fig. 1).

NGC 4013 is an Sbc Hubble type spiral galaxy at a distance
of 11.6 Mpc (Bottema 1995) and  a B-band luminosity M$_B$ = -18.1 mag. 
Photometric properties of the 2D stellar
distribution of this galaxy can be found in Bottema (1995), where
a disk with exponential scalelength of 2.3 kpc and a sech$^2$ scaleheight of
0.7 kpc is reported to fit the F-band data. Barnaby et al. (1993)
present bulge and disk decomposition of an H-band image, where
an exponential disk with scalelength 2.5 kpc and scaleheight 0.2 kpc
and a modified-Hubble type bulge with effective radius 0.24 kpc are
fitted. Observations of this galaxy in the 21 cm neutral hydrogen line
(Bottema 1995; 1996) show a highly warped gas disk, while the optical
image of this galaxy (also presented in Fig. 1) shows that the stellar distribution
in the disk is not warped, at least not to the degree that should
affect the axisymmetric modelling that we do [see also the optical
warp curves presented in Florido et al. (1991)].

\begin{table*}
\begin{center}
\caption[]{\it Observational data.}
\begin{tabular}{cccccccccc}
\hline
\multicolumn{1}{c}{Galaxy} &
\multicolumn{1}{c}{Pixel size} &
\multicolumn{1}{c}{Observing} &
\multicolumn{1}{c}{seeing ($\arcsec$)} &
\multicolumn{3}{c}{rms$^*$ (mag)}&
\multicolumn{3}{r}{Total exposure time (min)}\\
& $\arcsec$ ~~~~(pc) & nights & & B & V & I & B &~~~~~ V & I\\
\hline
NGC 4013 & 0.39 (22) & 20 Jun 95 & 1.3 & 0.04 & 0.02 & 0.02 &     &         &   \\
         & 0.39 (22) & 21 Jun 95 & 1.0 & 0.03 & 0.02 & 0.02 & 140 &~~~~~ 40 & 25\\
UGC 1082 & 0.39 (70) & 09 Sep 94 & 1.4 & 0.04 & 0.03 & 0.03 & 60  &~~~~~ 50 & 20\\
NGC 5529 & 0.39 (56) & 04 Aug 97 & 1.2 &      & 0.03 & 0.03 &     &~~~~~ 60 & 40\\
NGC 5907 & 0.75 (40) & 07 May 97 & 1.8 & 0.03 & 0.03 & 0.02 & 30  &~~~~~ 20 & 10\\
\hline
\end{tabular}
\\
$^*$ rms deviations from the least squares fit between catalogue magnitudes
and calibrated magnitudes for the standard stars.
\end{center}
\end{table*}

IC 2531 is an Sb Hubble type spiral galaxy at a distance of
22 Mpc (Shaw et al. 1990) and M$_B$ = -18.8 mag.
Two attempts at modelling IC 2531 have been reported (Wainscoat et al. 1989;
Just et al. 1996). The model used was along the lines
of KB87, but in a more simplified form, dealing with absorption
by dust and neglecting multiple scattering effects. Also, only a local
fit was made in several strips parallel to the disk's minor axis.
Furthermore, a bulge component was not included in the model.
Our work presents a global fit to the observed data using the full
radiative transfer model described in KB87 and more realistic
(bulge included) distributions.

UGC 1082 is an Sb Hubble type spiral galaxy with a redshift of 0.009353 $\pm$ 0.000017
(Giovanelli \& Haynes 1993).
Using the formulae given by Zombeck (1990), we derive a distance of
37 Mpc (assuming a value of H$_0$ = 75 km s$^{-1}$ Mpc$^{-1}$) for this object.
The B luminosity is M$_B$ = -18.4 mag.
Visual inspection of this galaxy (see Fig. 1) shows an edge-on configuration
of a relatively faint galactic disk and a bright bulge with an obvious
dust lane passing through the disk.

NGC 5529 is an Sc Hubble type spiral galaxy at a distance of 29.6 Mpc 
(Shaw et al. 1990) with M$_B$ = -19.6 mag.
Visual inspection of this system shows a galaxy which is not quite edge-on.
Despite this geometry, a dust lane is very well revealed and thus our
axisymmetric model, which is able to deal with any inclination angle,
can be applied.

NGC 5907 is a well studied Sc galaxy at a distance of 11 Mpc (van der Kruit
\& Searle 1982). It's B luminosity is M$_B$ = -19.1 mag. 
Several photometric studies have been reported 
about this galaxy and the stellar distribution has been obtained
in several bands. In van der Kruit \& Searle (1982), U, J and F band data were
fit to a truncated exponential disk in the R direction
and a sech$^2$ type function in the z direction. 
The inclination angle was reported to be 
$87\degr$,
while the scalelength and scaleheight were found to be 5.7 kpc and 0.43 kpc
respectively
in the J-band. Barnaby and Thronson (1992) used H-band data
to fit the stellar distribution of this galaxy. The disk was fitted with an 
exponential in the R direction with a scalelength of 4.0 kpc and a sech type 
function in the z direction with a scaleheight of 0.43 kpc. The bulge was
well described by a modified Hubble profile.
Morrison et al. (1994) obtained good quality R-band data and fitted the disk
with exponential (in both directions) functions with a scalelength of 4.81 kpc
and a scaleheight of 0.43 kpc. The inclination angle was reported
to be 86.7\degr. 

In Sect. 2 we describe the observations and the data reduction,
in Sect. 3 we briefly discuss the model and in Sect. 4 we present
the model results for the five galaxies. A discussion follows in Sect. 5
with some general conclusions.
A quick summary is presented in Sect. 6.

\section{Observations and data reduction}

Observations of the four galaxies NGC 4013, UGC 1082, NGC 5529 and 
NGC 5907 were made at Skinakas
observatory in Crete, using the 1.3 m telescope.
The detector that was used was a Thomson 1024 $\times$ 1024 CCD camera with 19$\mu$m pixels,
installed at the prime focus of the $f/7.7$ Ritchey-Cretien telescope.
This arrangement corresponds to 0.39 arcsec/pixel
and a field of view of $6.7\arcmin \times 6.7\arcmin$.
For the observations of NGC 5907 we used a focal reducer that
is installed on the telescope in order to increase the field of view.
This arrangement produced a larger field of view ($\sim 13\arcmin \times 13\arcmin$)
and a scale of 0.75 arcsec/pixel (using the same camera as above).
The chip was run with a gain of 3.33 $e^-$/ADU and showed a readout
noise of $5.6~ e^-$. Exposures were made through B, V, I broad band filters.
The B and V passbands are comparable to those of Johnson's photometric
system, while the I
passband is comparable to that of Cousin's photometric system. The effective
wavelength of the system (filter and camera) in each passband
is $0.443 \mu$m, $0.564 \mu$m and $0.809 \mu$m for the B, V and I bands
respectively.
In order to calibrate the images, we used standard stars and
a photometric procedure described in detail in Paper I.
In Table 1 we present some data concerning these observations.


\begin{table*}
\begin{center}
\caption[]{\it Global model fit parameters for NGC 4013.}
\begin{tabular}{llllllll}
\hline
\multicolumn{1}{c}{Parameter} &
\multicolumn{1}{c}{Units} &
\multicolumn{2}{c}{I band} &
\multicolumn{2}{c}{V band} &
\multicolumn{2}{c}{B band}\\
\hline
$I_s$ & mag/arcsec$^2$ &
$17.29$ & $\pm 0.05$ &
$19.08$ & $\pm 0.04$ &
$19.60$ & $\pm 0.06$ \\
$z_s$ & kpc & 0.20 & $\pm 0.01$ & 0.22 & $\pm 0.01 $ & 0.19 & $\pm 0.01$ \\
$h_s$ & kpc & 1.76 & $\pm 0.1 $& 1.94 & $\pm 0.1$ & 2.55 & $\pm 0.1$\\
$I_b$ & mag/arcsec$^2$ &
$10.49$ & $\pm 0.13$ &
$11.73$ & $\pm 0.11$ &
$12.60$ & $\pm 0.14$ \\
$R_e$ & kpc & 1.61 & $\pm 0.06 $& 1.42 & $\pm 0.04 $& 1.46 & $\pm 0.05 $\\
$b/a$ & -- & 0.42 & $ \pm 0.01 $ &0.44 &
$ \pm 0.01 $& 0.41 & $ \pm 0.01 $ \\
$\tau_{\lambda}^f$ & -- & 0.48 & $\pm 0.01 $ & 0.67 &
$\pm 0.01 $& 0.89 & $\pm 0.01$ \\
$z_d$ & kpc & 0.11 & $\pm 0.01 $ & 0.13 & $\pm 0.01 $& 0.13 & $\pm 0.01 $\\
$h_d$ & kpc & 2.60 & $\pm 0.1$ & 2.45 & $\pm 0.1$ & 2.62 & $\pm 0.3$ \\
$\theta$ & degrees & 89.6 & $\pm 0.1$ &89.7 &$\pm 0.1$ & 89.9&$ \pm 0.1$\\
\hline
\end{tabular}
\end{center}
\end{table*}

\begin{table*}[t]
\begin{center}
\caption[]{\it Global model fit parameters for IC 2531.}
\begin{tabular}{llllllllllll}
\hline
\multicolumn{1}{c}{Parameter} &
\multicolumn{1}{c}{Units} &
\multicolumn{2}{c}{K band} &
\multicolumn{2}{c}{J band} &
\multicolumn{2}{c}{I band} &
\multicolumn{2}{c}{V band} &
\multicolumn{2}{c}{B band}\\
\hline
$I_s$ & mag/arcsec$^2$ &
$16.49$ & $\pm 0.12$ &
$17.45$ & $\pm 0.35$ &
$18.24$ & $\pm 0.04$ &
$19.48$ & $\pm 0.05$ &
$20.34$ & $\pm 0.07$ \\
$z_s$ & kpc &0.45 & $\pm 0.02$ & 0.44 & $\pm 0.10 $ & 0.43 & $\pm 0.01$
& 0.40 & $\pm 0.01$ & 0.43 & $\pm 0.01 $\\
$h_s$ & kpc &5.04 & $\pm 0.1 $& 4.96& $\pm 0.3$ & 5.05& $\pm 0.1$
& 5.22& $\pm 0.1 $& 6.78 & $\pm 0.1$\\
$I_b$ & mag/arcsec$^2$ &
$10.39$ & $\pm 0.37$ &
$10.97$ & $\pm 0.40$ &
$11.00$ & $\pm 0.20$ &
$11.85$ & $\pm 0.20$ &
$13.40$ & $\pm 0.30$ \\
$R_e$ & kpc & 2.00 & $\pm 0.24 $& 1.97 & $\pm 0.10 $& 1.57 & $\pm 0.08 $
& 1.23 & $\pm 0.08 $& 1.96 & $\pm 0.15 $\\
$b/a$ & -- & 0.68 & $ \pm 0.03 $ &0.69 &
$ \pm 0.02 $& 0.65& $ \pm 0.02 $ & 0.65& $ \pm 0.02 $
& 0.63& $ \pm 0.02 $\\
$\tau_{\lambda}^f$ & -- & 0.02& $\pm 0.01 $ & 0.06&
$\pm 0.02 $& 0.22& $\pm 0.01$ & 0.30& $\pm 0.01 $& 0.40& $\pm 0.01 $\\
$z_d$ & kpc &0.22 & $\pm 0.03 $ &0.20 & $\pm 0.07 $& 0.21& $\pm 0.01 $
& 0.23& $\pm 0.01 $ &0.27 & $\pm 0.01 $\\
$h_d$ & kpc &8.00 & $\pm 0.3$ & 8.08 & $\pm 0.6$ &8.43 & $\pm 0.2$
& 8.18& $\pm 0.2$ & 8.88& $\pm 0.3$\\
$\theta$ & degrees & 89.6 & $\pm 0.2$ & 89.7 & $\pm 0.2$ & 89.6&$ \pm 0.2$
& 89.6 & $\pm 0.2$ &89.6 &$\pm 0.2$\\
\hline
\end{tabular}
\end{center}
\end{table*}

\begin{table*}
\begin{center}
\caption[]{\it Global model fit parameters for UGC 1082.}
\begin{tabular}{llllllll}
\hline
\multicolumn{1}{c}{Parameter} &
\multicolumn{1}{c}{Units} &
\multicolumn{2}{c}{I band} &
\multicolumn{2}{c}{V band} &
\multicolumn{2}{c}{B band}\\
\hline
$I_s$ & mag/arcsec$^2$ &
$19.02$ & $\pm 0.11$ &
$20.50$ & $\pm 0.07$ &
$21.29$ & $\pm 0.09$ \\
$z_s$ & kpc & 0.44& $\pm 0.02$ & 0.47& $\pm 0.02 $ & 0.44& $\pm 0.03$ \\
$h_s$ & kpc &4.13 & $\pm 0.2 $&4.43 &$\pm 0.1$ & 4.57& $\pm 0.2$\\
$I_b$ & mag/arcsec$^2$ &
$9.42$ & $\pm 0.08$ &
$10.98$ & $\pm 0.10$ &
$12.17$ & $\pm 0.09$ \\
$R_e$ & kpc & 1.34 & $\pm 0.12 $& 1.30 & $\pm 0.14 $& 1.44& $\pm 0.16 $\\
$b/a$ & -- & 0.59 & $ \pm 0.01 $ &0.64 &
$ \pm 0.01 $& 0.63 & $ \pm 0.01 $ \\
$\tau_{\lambda}^f$ & -- & 0.17 & $\pm 0.01 $ & 0.27 &
$\pm 0.02 $& 0.34 & $\pm 0.02$ \\
$z_d$ & kpc & 0.29 & $\pm 0.02 $ & 0.27 & $\pm 0.02 $& 0.27 & $\pm 0.03 $\\
$h_d$ & kpc & 5.47 & $\pm 0.4$ & 5.78 & $\pm 0.3$ & 5.55 & $\pm 0.4$ \\
$\theta$ & degrees & 89.7 & $\pm 0.1$ &89.7 &$\pm 0.1$ & 89.7&$ \pm 0.1$\\
\hline
\end{tabular}
\end{center}
\end{table*}

\begin{table*}
\begin{center}
\caption[]{\it Global model fit parameters for NGC 5529.}
\begin{tabular}{llllll}
\hline
\multicolumn{1}{c}{Parameter} &
\multicolumn{1}{c}{Units} &
\multicolumn{2}{c}{I band} &
\multicolumn{2}{c}{V band}\\
\hline
$I_s$ & mag/arcsec$^2$ &
$17.13$ & $\pm 0.05$ &
$18.42$ & $\pm 0.04$ \\
$z_s$ & kpc & 0.42 & $\pm 0.01$ & 0.43 & $\pm 0.01 $ \\
$h_s$ & kpc & 4.30 & $\pm 0.1 $& 4.59& $\pm 0.1$ \\
$I_b$ & mag/arcsec$^2$ &
$10.98$ & $\pm 0.06$ &
$12.25$ & $\pm 0.07$ \\
$R_e$ & kpc & 1.95& $\pm 0.01 $& 1.95 & $\pm 0.01 $\\
$b/a$ & -- & 0.59 & $ \pm 0.01 $ &0.58 &
$ \pm 0.01 $ \\
$\tau_{\lambda}^f$ & -- & 0.42 & $\pm 0.01 $ & 0.65 &
$\pm 0.01 $\\
$z_d$ & kpc & 0.37 & $\pm 0.01 $ & 0.32 & $\pm 0.01 $\\
$h_d$ & kpc & 7.70 & $\pm 0.2$ & 7.10 & $\pm 0.1$ \\
$\theta$ & degrees & 87.2 & $\pm 0.1$ &87.4 &$\pm 0.1$\\
\hline
\end{tabular}
\end{center}
\end{table*}

\begin{table*}
\begin{center}
\caption[]{\it Global model fit parameters for NGC 5907.}
\begin{tabular}{llllllll}
\hline
\multicolumn{1}{c}{Parameter} &
\multicolumn{1}{c}{Units} &
\multicolumn{2}{c}{I band} &
\multicolumn{2}{c}{V band} &
\multicolumn{2}{c}{B band}\\
\hline
$I_s$ & mag/arcsec$^2$ &
$17.49$ & $\pm 0.02$ &
$18.99$ & $\pm 0.02$ &
$19.86$ & $\pm 0.03$ \\
$z_s$ & kpc & 0.32 & $\pm 0.01$ & 0.34 & $\pm 0.01 $ & 0.34 & $\pm 0.01$ \\
$h_s$ & kpc & 3.86 & $\pm 0.1 $& 4.91 &$\pm 0.1$ & 5.02 & $\pm 0.1$\\
$I_b$ & mag/arcsec$^2$ &
$9.38$ & $\pm 0.04$ &
$10.80$ & $\pm 0.04$ &
$11.88$ & $\pm 0.05$ \\
$R_e$ & kpc & 1.02 & $\pm 0.01 $& 1.01 & $\pm 0.01 $& 0.93& $\pm 0.01 $\\
$b/a$ & -- & 0.34 & $ \pm 0.01 $ &0.34 &
$ \pm 0.01 $& 0.45 & $ \pm 0.01 $ \\
$\tau_{\lambda}^f$ & -- & 0.30 & $\pm 0.01 $ & 0.49 &
$\pm 0.01 $ & 0.55 & $\pm 0.01$ \\
$z_d$ & kpc & 0.10 & $\pm 0.01 $ & 0.11 & $\pm 0.01 $& 0.13 & $\pm 0.01 $\\
$h_d$ & kpc & 5.30 & $\pm 0.3$ & 5.29 & $\pm 0.2$ & 5.30 & $\pm 0.2$ \\
$\theta$ & degrees & 87.2 & $\pm 0.1$ &87.2 &$\pm 0.2$ & 87.0&$ \pm 0.2$\\
\hline
\end{tabular}
\end{center}
\end{table*}

\begin{table*}
\begin{center}
\caption[]{\it Global model fit parameters for NGC 891.}
\begin{tabular}{llllllllllll}
\hline
\multicolumn{1}{c}{Parameter} &
\multicolumn{1}{c}{Units} &
\multicolumn{2}{c}{K band} &
\multicolumn{2}{c}{J band} &
\multicolumn{2}{c}{I band} &
\multicolumn{2}{c}{V band} &
\multicolumn{2}{c}{B band}\\
\hline
$I_s$ & mag/arcsec$^2$ &
$15.41$ & $\pm 0.13$ &
$16.01$ & $\pm 0.11$ &
$17.43$ & $\pm 0.03$ &
$18.89$ & $\pm 0.05$ &
$19.81$ & $\pm 0.06$ \\
$z_s$ & kpc & 0.34 & $\pm 0.01$ & 0.43 & $\pm 0.01 $ & 0.38 & $\pm 0.01$
& 0.42& $\pm 0.01$ &0.43 & $\pm 0.01 $\\
$h_s$ & kpc & 3.87& $\pm 0.1 $& 3.86& $\pm 0.1$ & 4.93& $\pm 0.1$
& 5.48& $\pm 0.2 $& 5.67& $\pm 0.2$\\
$I_b$ & mag/arcsec$^2$ &
$8.32$ & $\pm 0.54$ &
$9.35$ & $\pm 0.16$ &
$10.38$ & $\pm 0.11$ &
$10.97$ & $\pm 0.22$ &
$11.42$ & $\pm 0.32$ \\
$R_e$ & kpc & 0.86 & $\pm 0.05 $& 0.87 & $\pm 0.05 $& 1.97 & $\pm 0.06 $
& 1.51 & $\pm 0.09 $& 1.12 & $\pm 0.09 $\\
$b/a$ & -- & 0.76 & $ \pm 0.04 $ &0.71 &
$ \pm 0.08 $& 0.54 & $ \pm 0.01 $ & 0.54 & $ \pm 0.01 $
& 0.60 & $ \pm 0.02 $\\
$\tau_{\lambda}^f$ & -- & 0.09 & $\pm 0.01 $ & 0.21 &
$\pm 0.01 $& 0.52 & $\pm 0.01$ & 0.85 & $\pm 0.01 $& 1.00 & $\pm 0.01 $\\
$z_d$ & kpc & 0.25 & $\pm 0.03 $ & 0.25 & $\pm 0.02 $& 0.24 & $\pm 0.01 $
& 0.29 & $\pm 0.01 $ & 0.31 & $\pm 0.02 $\\
$h_d$ & kpc & 8.33 & $\pm 0.4$ & 8.32 & $\pm 0.4$ & 7.54 & $\pm 0.1$
& 7.68 & $\pm 0.2$ & 8.01 & $\pm 0.3$\\
$\theta$ & degrees & 89.6 & $\pm 0.1$ &89.6 &$\pm 0.1$ & 89.7&$ \pm 0.1$
& 89.8 & $\pm 0.2$ &89.8 &$\pm 0.2$\\
\hline
\end{tabular}
\end{center}
\end{table*}
 
The optical observations of IC 2531 were made in March, 1991 with the 1m
Australian National University (ANU) telescope at the Siding Spring
Observatory (SSO).  The CCD system was equipped with an EEV CCD of $576 \times
380$
giving a pixel size of 0.56 arcsec at the f/8 Cassegrain focus.  Graham's
(1982) standard stars were observed for calibration.
Observations were made through B, V and I broad band filters with exposure
times of 1000, 600, and 500 seconds respectively.
The seeing was estimated to be $ 2 \arcsec$.
The near-infrared observations of the same galaxy were made in February, 1995 with 
the 2.3 m ANU
telescope at SSO using CASPIR (Cryogenic Array Spectrometer/Imager).  The
instrument was installed at the f/18 Cassegrain focus and was equipped with an
SBRC InSb $ 256 \times 256$
detector giving a pixel size of 0.5 arcsec for its fast
camera mode.  The standard $J$ filter and a modified $K$ filter ($Kn$), whose
long wavelength edge is tailored to exclude much of the thermal emission,
were used for the imaging observations.  The observation sequence included as
many sky exposures as the galaxy in order to follow the fast changing sky
emissions.   Several standard stars were observed during the same night.
 
\section{Model}
The stellar emissivity (luminosity per unit volume)
that we use consists of an
exponential (in both radial and vertical directions) disk and
a bulge described by the R$^{1/4}$ law, namely
\begin{displaymath}
L(R,z) = L_s \exp \left( - \frac{R}{h_s} - \frac{|z|}{z_s} \right)
\end{displaymath}
\begin{equation}
~~~~~~+ L_b \exp (-7.67 B^{1/4}) B^{-7/8},
\end{equation}
with $h_s$ and $z_s$ being the scalelength and scaleheight of the
disk and
\begin{equation}
B = \frac{\sqrt{R^2 + z^2 (b/a)^2}}{R_e} ,
\end{equation}
with $R_e$ being the effective radius of the bulge and $a$ and $b$
the semi-major and semi-minor axis respectively.
Here $L_s$ and $L_b$ are the normalization constants for the stellar
emissivity of the disk and the bulge respectively. The relations
\begin{equation}
I_s = 2L_sh_s ,
\end{equation}
and
\begin{equation}
I_b = 5.12L_bR_e ,
\end{equation}
give the central value for the surface brightness
of the disk and the bulge respectively, if the model galaxy
is seen edge-on and there is no dust.

For the extinction coefficient we use a double exponential law, namely
\begin{equation}
\kappa_{\lambda}(R,z) = \kappa_{\lambda}
\exp \left(- \frac{R}{h_d} - \frac{|z|}{z_d}
\right) ,
\end{equation}
where $\kappa_{\lambda}$ is the extinction coefficient at
wavelength $\lambda$ at the center of the disk and
$h_d$ and $z_d$ are the scalelength and scaleheight respectively
of the dust. The central optical depth of the model galaxy seen face-on
is
\begin{equation}
\tau^f_{\lambda} = 2\kappa_{\lambda}z_d .
\end{equation}

For a more detailed description of all the parameters, the reader
is referred to Sect. 4 of Paper I.
 
\begin{figure}[!ht]
\epsfysize=9.5cm
\epsfbox [86 254 520 739]{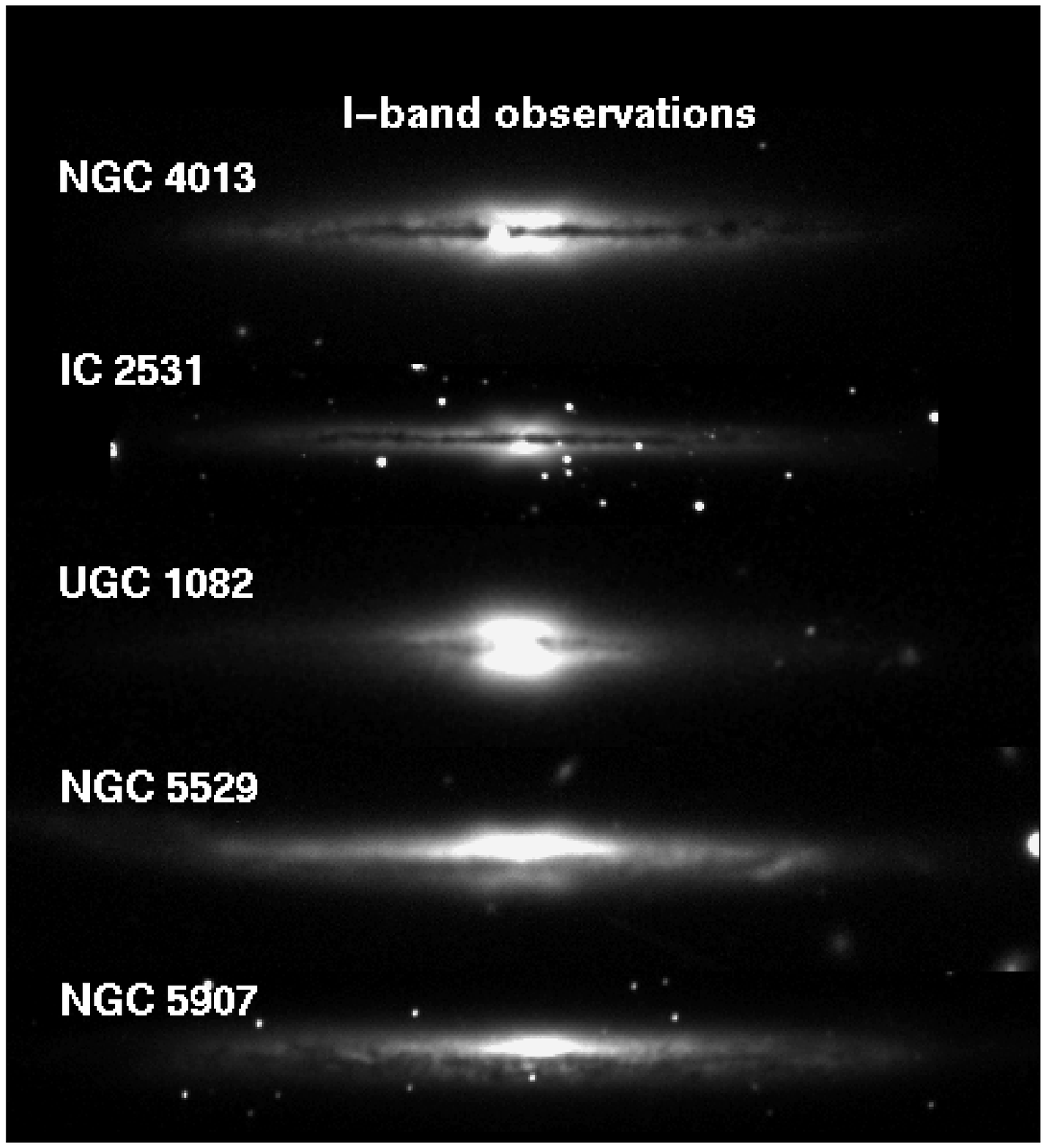}
\caption{I-band observations of the galaxies
NGC 4013, IC 2531, UGC 1082, NGC 5529 and NGC 5907
(top to bottom).}
\end{figure}
\begin{figure}
\epsfysize=9.5cm
\epsfbox [86 254 520 739]{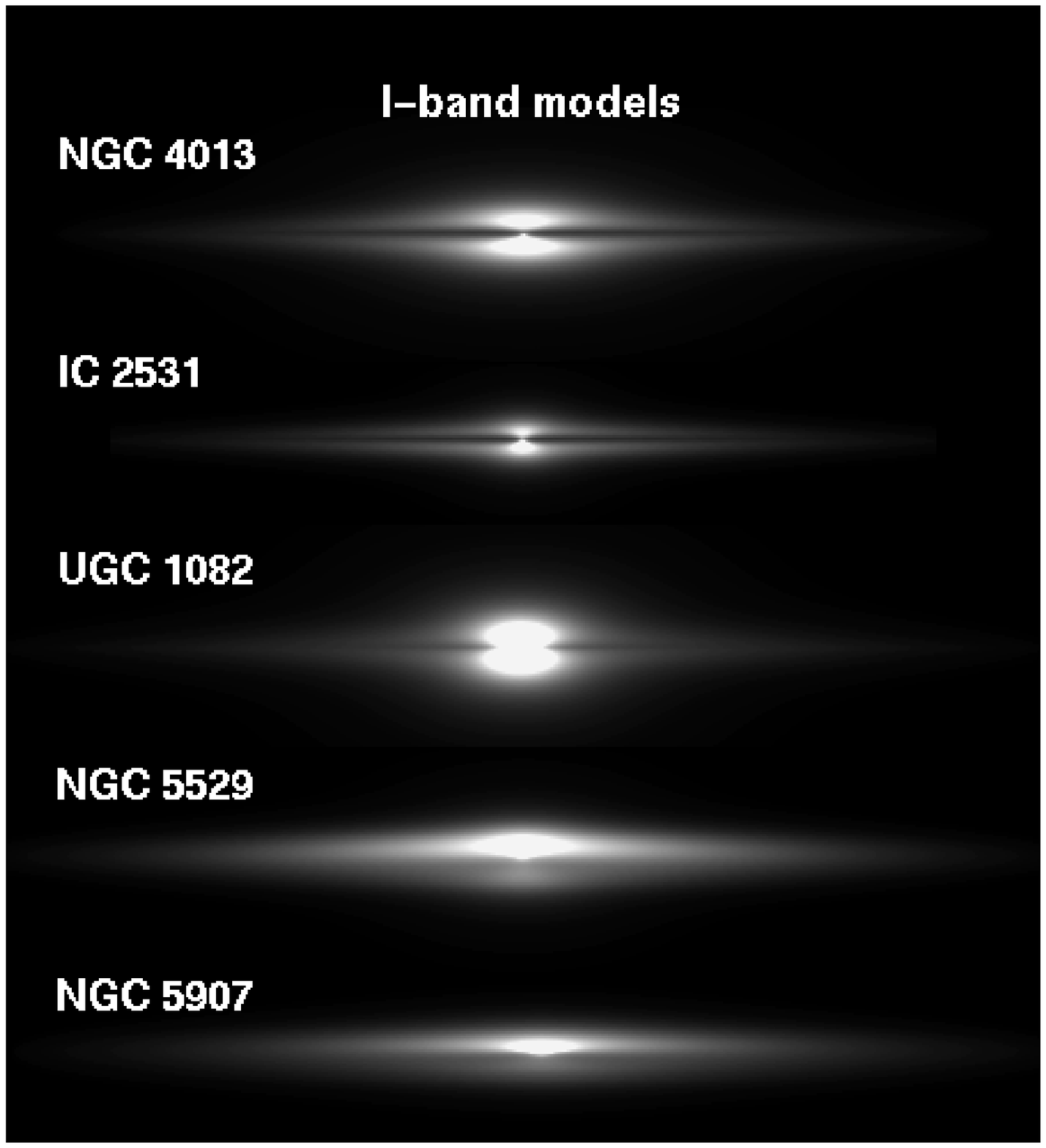}
\caption{I-band models of the galaxies
NGC 4013, IC 2531, UGC 1082, NGC 5529 and NGC 5907
(top to bottom).}
\end{figure}
The radiative transfer
model that we have used is that described in KB87 (see also Paper I).
For computational reasons, as in Papers I \& II,
the radiative transfer was performed inside a cylinder with radius
$R_{max} = 3 \max(h_s,h_d)$ and half height of $6 \max(z_s,R_e)$,
so that practically all the galactic light as well as the dust
are included.
A Henyey-Greenstein phase function has
been used for the scattering of the dust (Henyey \& Greenstein 1941).
The values for the anisotropy parameter $g$ and the albedo $\omega$
have been taken from Bruzual et al. (1988).
Our task has been to find those values of the parameters in Eqs. (1) - (6)
which create images of the model galaxies as close as possible to
the images of the observed galaxies.
For the model fitting techniques that we use, the reader is referred
to Papers I and II.

\section{Results}
In Tables 2 through 6, we give the parameters of the best fit models
to the observed data
along with their $95\%$ confidence intervals
for NGC 4013, IC 2531, UGC 1082, NGC 5529 and NGC 5907
respectively (for Table 7 describing NGC 891 see Sect. 5).
Using the values of the parameters that
describe the stellar and dust distributions of the galaxies,
given in the above tables,
we have created model images and have compared them with the
observed images
(see Figs. 1 and 2). In Fig. 1 the I-band observations of all five
galaxies are presented (NGC 4013, IC 2531, UGC 1082, NGC 5529 and NGC 5907
from top to bottom). In Fig. 2 we show the model images
of these galaxies in the I-band with the same scale
and sequence as in Fig. 1 so that a direct comparison can be made
between the two figures.
In Figs. 3-7 we give a more detailed comparison between model and
observation for each galaxy and each filter by showing vertical cuts
along the minor axis. For this demonstration we use the `folded'
(around the minor axis) and photometrically averaged images of the galaxies,
which are also the images that were used for the model fit (see Paper I).
In each plot, the horizontal axis represents the offset (in kpc)
along the vertical direction, with zero lying on the major axis
of the disk. The vertical axis gives the surface
brightness (in mag/arcsec$^2$). Real data are indicated by stars, while
the model is shown as a solid line for each profile. The six
profiles in each plot are vertical cuts at six different
distances along the major axis.
The cuts are at distances $0, 0.5h_s^I, 1h_s^I, 1.5h_s^I, 2h_s^I$
and $2.5h_s^I$ and are plotted from bottom to top, with $h_s^I$ being the
scalelength of the stars as derived from the I-band modelling.
The magnitude scale corresponds to the lower profile of each set, the other
profiles are shifted upwards (in brightness) by 2, 4, 6, 8 and 10
magnitudes respectively. Profiles between 0 and $1.5h_s^I$ correspond
to brightness averaged over $4\arcsec$ ($7.5\arcsec$ for NGC 5907)
parallel to the major axis, while profiles at $2h_s^I$ and $2.5h_s^I$
are averaged over $8\arcsec$ ($15\arcsec$ for NGC 5907) along the
same direction. This was done to reduce as much as possible
the local clumpiness that may exist in particular areas of the galaxy.
In these plots, only the data above the limiting
sigma level ($\sim 3$ sigma of the local sky)
that were used for the fit are plotted,
while the foreground stars have been removed.

\begin{figure*}[!ht]
\epsfysize=6cm
\epsfbox [23 27 565 269]{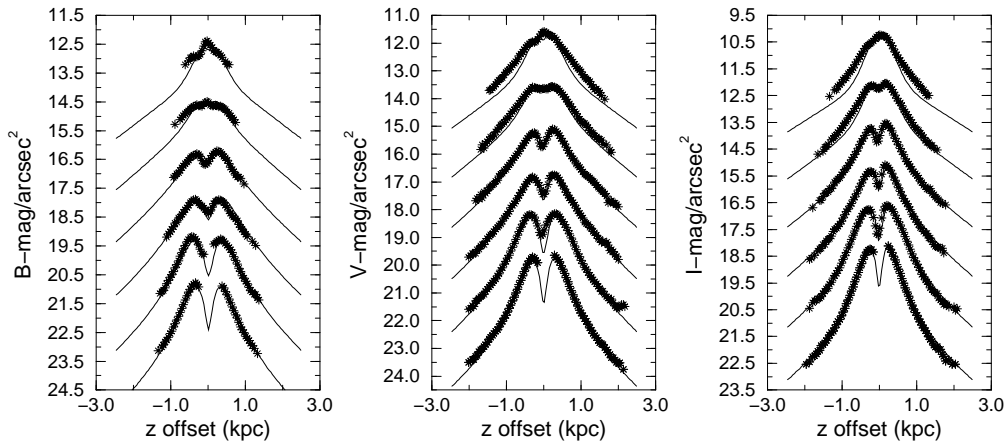}
\caption{Vertical surface brightness profiles for NGC 4013 in the
B (left), V (middle), I (right) bands. The six profiles in each plot
are vertical cuts of the galactic plane at distances of
$0, 0.5h_s^I, 1h_s^I, 1.5h_s^I, 2h_s^I$
and $2.5h_s^I$ along the major axis as plotted from bottom to top
respectively. Stars represent averaged profiles of the `folded'
image of the observation, while solid lines give the 
model. From bottom to top, the magnitude scale is shifted upwards
(in brightness) by 0, 2, 4, 6, 8 and 10 magnitudes respectively. See the
text for a more detailed description.}
\end{figure*}

\begin{figure*}
\epsfysize=12cm
\epsfbox [15 15 558 500]{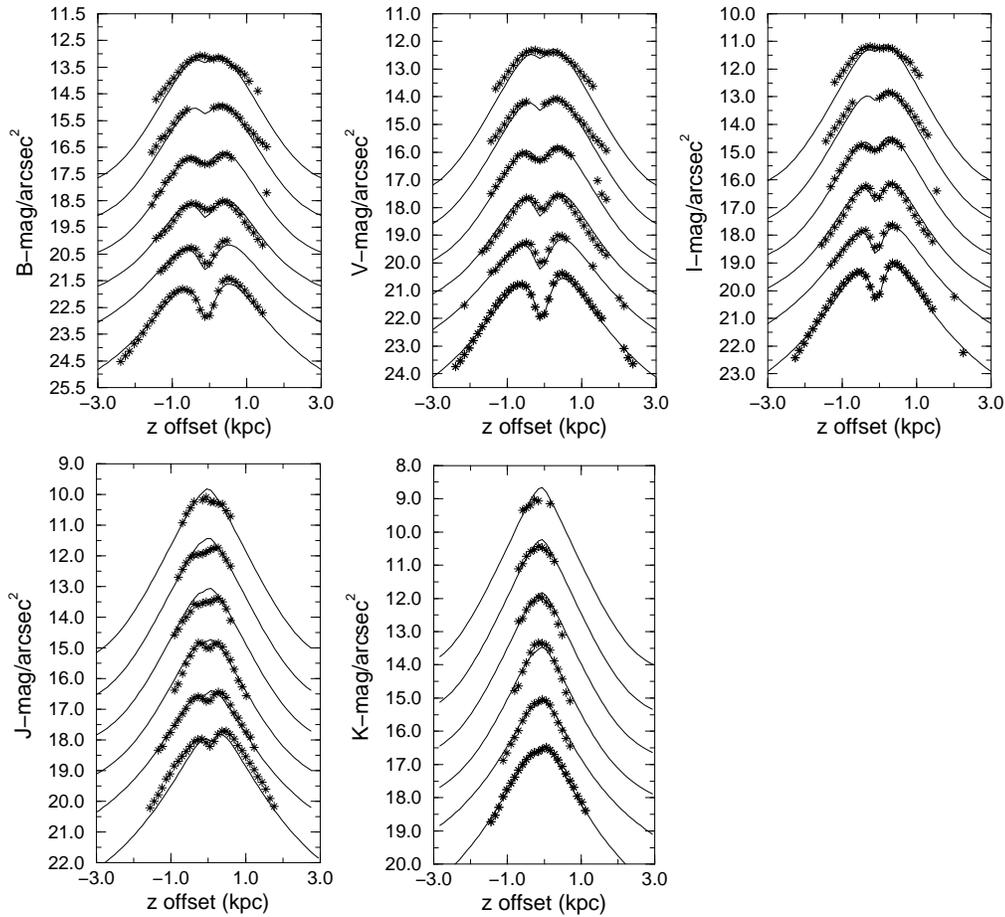}
\caption{Same as in Fig. 3, but for IC 2531.
In addition, J and K band cuts are shown.}
\end{figure*}

\begin{figure*}
\epsfysize=6cm
\epsfbox [23 22 565 274]{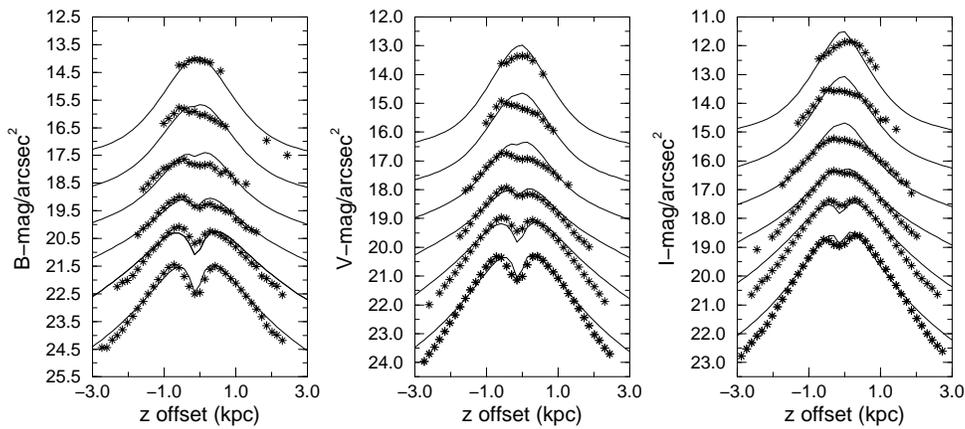}
\caption{Same as in Fig. 3, but for UGC 1082.}
\end{figure*}

\begin{figure*}
\epsfysize=6cm
\epsfbox [76 175 450 423]{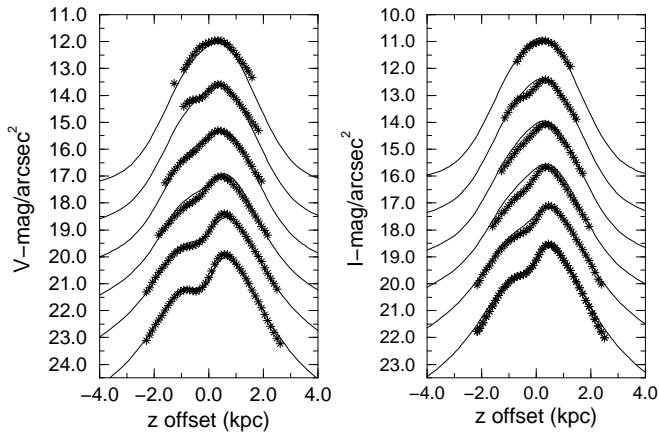}
\caption{Same as in Fig. 3, but for NGC 5529.}
\end{figure*}

\begin{figure*}
\epsfysize=6cm
\epsfbox [23 26 565 274]{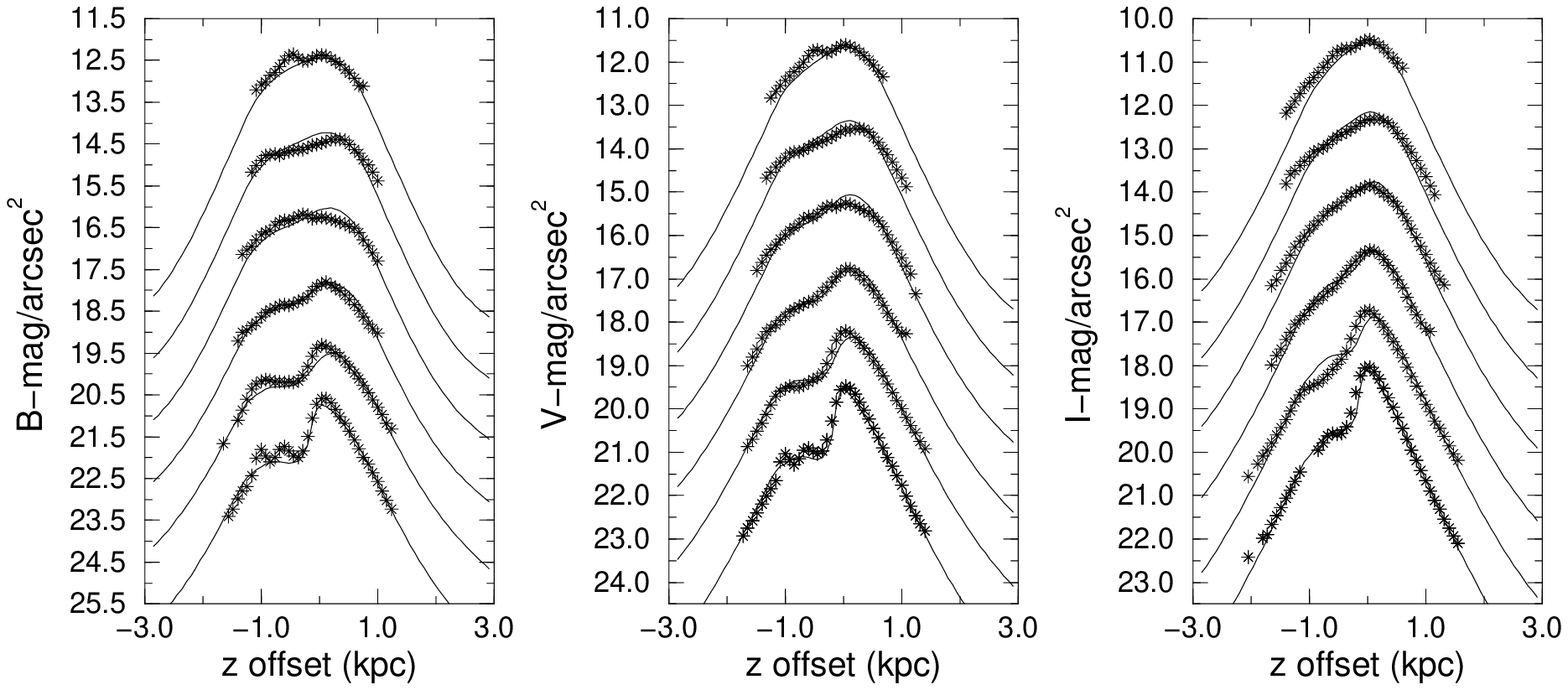}
\caption{Same as in Fig. 3, but for NGC 5907.}
\end{figure*}

One can clearly see a good agreement between model and observation
despite the small deviations in some places. It is worth remembering
here
that a {\it global} fit was done to the observed
galaxy's image and the derived set of parameters (Tables 2 - 6)
gives the best galaxy image as a whole.

\begin{table}
\begin{center}
\caption[]{\it Percentage coverage of the
galaxy's image when the absolute value of the residuals
is below a certain value.}
\begin{tabular}{llllll}
\hline
\multicolumn{1}{c}{Galaxy} &
\multicolumn{1}{c}{$<10\%$} &
\multicolumn{1}{c}{$<20\%$} &
\multicolumn{1}{c}{$<30\%$} &
\multicolumn{1}{c}{$<40\%$} &
\multicolumn{1}{c}{$<50\%$}\\
\hline
NGC 4013 & $46.6\%$ & $73.2\%$ & $87.0\%$ & $94.3\%$ & $97.8\%$ \\
IC 2531  & $35.8\%$ & $66.0\%$ & $83.4\%$ & $94.4\%$ & $99.5\%$ \\
UGC 1082 & $38.0\%$ & $65.0\%$ & $82.2\%$ & $91.1\%$ & $95.0\%$ \\
NGC 5529 & $41.7\%$ & $73.9\%$ & $87.6\%$ & $94.1\%$ & $97.5\%$ \\
NGC 5907 & $38.5\%$ & $66.2\%$ & $83.0\%$ & $92.7\%$ & $97.4\%$ \\
\hline
\end{tabular}
\end{center}
\end{table}

Subtracting the model image from the observation we obtain residual
maps that give the distribution of  the absolute value of 
the residuals throughout the 
galaxy's image (see Papers I \& II). From these maps we can 
derive the percentage coverage of the galaxy's image with
residuals below a certain value.
We do so in Table 8, where the percentage coverage of the
residuals below $10\%, 20\%, 30\%, 40\%$ and $50\%$ is given for the
I-band data. Approximately the same picture holds for the other bands
(see Papers I \& II). These numbers indicate the ability of our model
to describe quite accurately the observed surface brightness of the 
galaxies that we modelled using relatively simple and smooth
three dimensional
distributions of stars and dust compared to the real ones with all the
small scale clumpiness and spiral structure that they may have.

\subsection{Dust mass calculations}
Having found the parameters that best describe the dust distribution
in each galaxy, we are able to have a good estimate of the total dust
mass of each galaxy (see Paper I). In what follows we do these
calculations and also compare with the total gas mass reported in the 
literature as well as with the dust mass calculated using the 
IRAS fluxes (wherever measurements are available).

For NGC 4013 the mass of the atomic hydrogen is reported to be
$1.3 \times 10^9 M_{\sun}$ (Bottema 1995), while the mass of the molecular
hydrogen is $8.9 \times 10^8 M_{\sun}$ (Gomez \& Garcia 1997), giving
a total of $M_g = 2.2 \times 10^9 M_{\sun}$ for the gas mass.
Using the parameters derived from our model, the dust mass was calculated
(see Sect. 5.3 of Paper I) to be $M_d = 4.5 \times 10^6 M_{\sun}$.
This implies a gas-to-dust mass ratio of 490.
Using the IRAS $60 \mu m$ and $100 \mu m$ fluxes for this galaxy
(Moshir et al. 1990) and Eq. (4) of Devereux \& Young (1990), the dust mass
is calculated to be $1.9 \times 10^6 M_{\sun}$, giving
a gas-to-dust mass ratio of 1200. 

In the case of IC 2531 the 21 cm line flux was measured to be
S$_{H_I} = 40.3$ Jy km s$^{-1}$ (Huchtmeier \& Richter 1989)
and thus the atomic hydrogen mass was calculated using Eq. (2) of 
Devereux \& Young (1990) and is found to be 
$M(H_I) = 4.6 \times 10^9 M_{\sun}$.
We have not been able to locate a measurement for the 2.6 mm CO line
for this galaxy in order to calculate the molecular hydrogen
mass $M(H_2)$, but a crude approximation is to assume the same mass
as that of the atomic hydrogen (see Sect. 5.3 of Paper I). Thus, the
total gas mass is approximately $M_g = 9.2 \times 10^9 M_{\sun}$.
The dust mass for this galaxy as calculated from our model parameters is
$M_d = 2.2 \times 10^7 M_{\sun}$, setting the gas-to-dust mass ratio to
420.

For UGC 1082 the 21 cm line flux has been reported to be 
S$_{H_I} = 5.47 $ Jy km s$^{-1}$ (Giovanelli \& Haynes 1993).
This gives an atomic hydrogen mass of $M(H_I) = 1.8 \times 10^9 M_{\sun}$.
In absence of a 2.6 mm CO line measurement in order to calculate the
molecular hydrogen mass we approximate it to be 
the same as that of the atomic hydrogen.
The total gas mass is then $M_g = 3.6 \times 10^9 M_{\sun}$.
Using the IRAS $60 \mu m$ and $100 \mu m$ fluxes for this galaxy
(Moshir et al. 1990) and Eq. (4) of Devereux \& Young (1990), the dust mass
is found to be $1.9 \times 10^6 M_{\sun}$. This implies a gas-to-dust
mass ratio of 1900. However, using the parameters derived from our
model, we calculate a dust mass of $M_d = 9.9 \times 10^6 M_{\sun}$ and the
gas-to-dust mass ratio then becomes 360.

Adopting a 21 cm line flux of 40.8 Jy km s$^{-1}$ (Huchtmeier \& Richter 1989)
for NGC 5529,
we derive an atomic hydrogen mass of $M(H_I) = 8.4 \times 10^9 M_{\sun}$.
Doubling it, in the absence of molecular hydrogen mass information,
we get a total gas mass of $M_g = 1.7 \times 10^{10} M_{\sun}$.
The $60 \mu m$ and $100 \mu m$ IRAS fluxes (Moshir et al. 1990)
give a dust mass of $5.7 \times 10^6 M_{\sun}$ which implies
a gas-to-dust mass ratio of 3000. With our model calculations
we find a dust mass of $3.6 \times 10^7 M_{\sun}$, giving a 
gas-to-dust mass ratio of 470.

For NGC 5907 exact gas mass calculations are presented in
Dumke et al. (1997). 
The total gas mass is reported to be $M_g = 7.8 \times 10^9 M_{\sun}$.
The IRAS $60 \mu m$ and $100 \mu m$ fluxes (Moshir et al. 1990)
give a dust mass of $3.6 \times 10^6 M_{\sun}$, while
with our model parameters we derive a dust mass of
$1.5 \times 10^7 M_{\sun}$. The gas-to-dust mass ratio
then becomes 2200 using the IRAS measurements and
520 using our model's parameters.

\subsection{Relative extinction values}
Adopting a mean scaleheight for the dust from all bands (see Sect. 5)
and using the values derived for the 
face-on central optical depth ($\tau_{\lambda}^f = 2 \kappa_{\lambda} z_d$)
in each filter, we can derive the extinction coefficient $\kappa_{\lambda}$
in each band. We have found that
$\kappa_B/\kappa_V = 1.33$ and $\kappa_I/\kappa_V = 0.72$ for NGC 4013,
$\kappa_B/\kappa_V = 1.33$, $\kappa_I/\kappa_V = 0.74$,
$\kappa_J/\kappa_V = 0.20$ anad $\kappa_K/\kappa_V = 0.06$ for IC 2531,
$\kappa_B/\kappa_V = 1.26$ and $\kappa_I/\kappa_V = 0.62$ for UGC 1082,
$\kappa_I/\kappa_V = 0.65$ for NGC 5529 and
$\kappa_B/\kappa_V = 1.12$ and $\kappa_I/\kappa_V = 0.61$ for NGC 5907.
These ratios are identical to the ratios of the
extinction values $A_{\lambda}/A_V$.

\begin{figure}
\epsfxsize=8.9cm
\epsfbox [70 122 488 427]{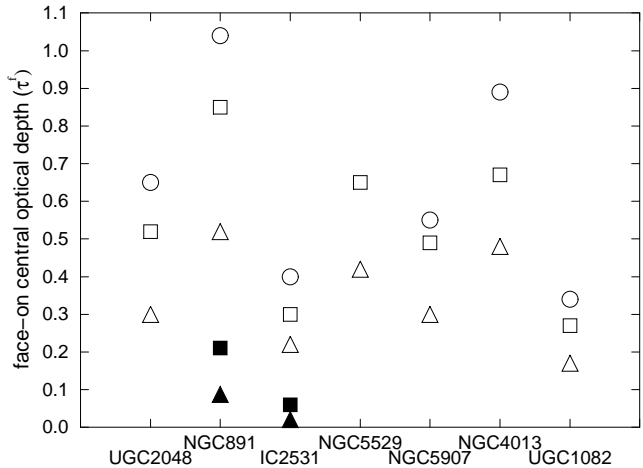}
\caption{Values of the face-on central optical depth in all the
bands that were modelled for each galaxy. 
Different symbols are for different bands.
Open circles correspond to the B-band, open squares to the V-band,
open triangles to the I-band, solid squares to the J-band and solid triangles
to the K-band. For UGC 2048 the values are taken from Paper I.}
\end{figure}
\begin{figure}
\epsfxsize=8.9cm
\epsfbox [93 49 483 389]{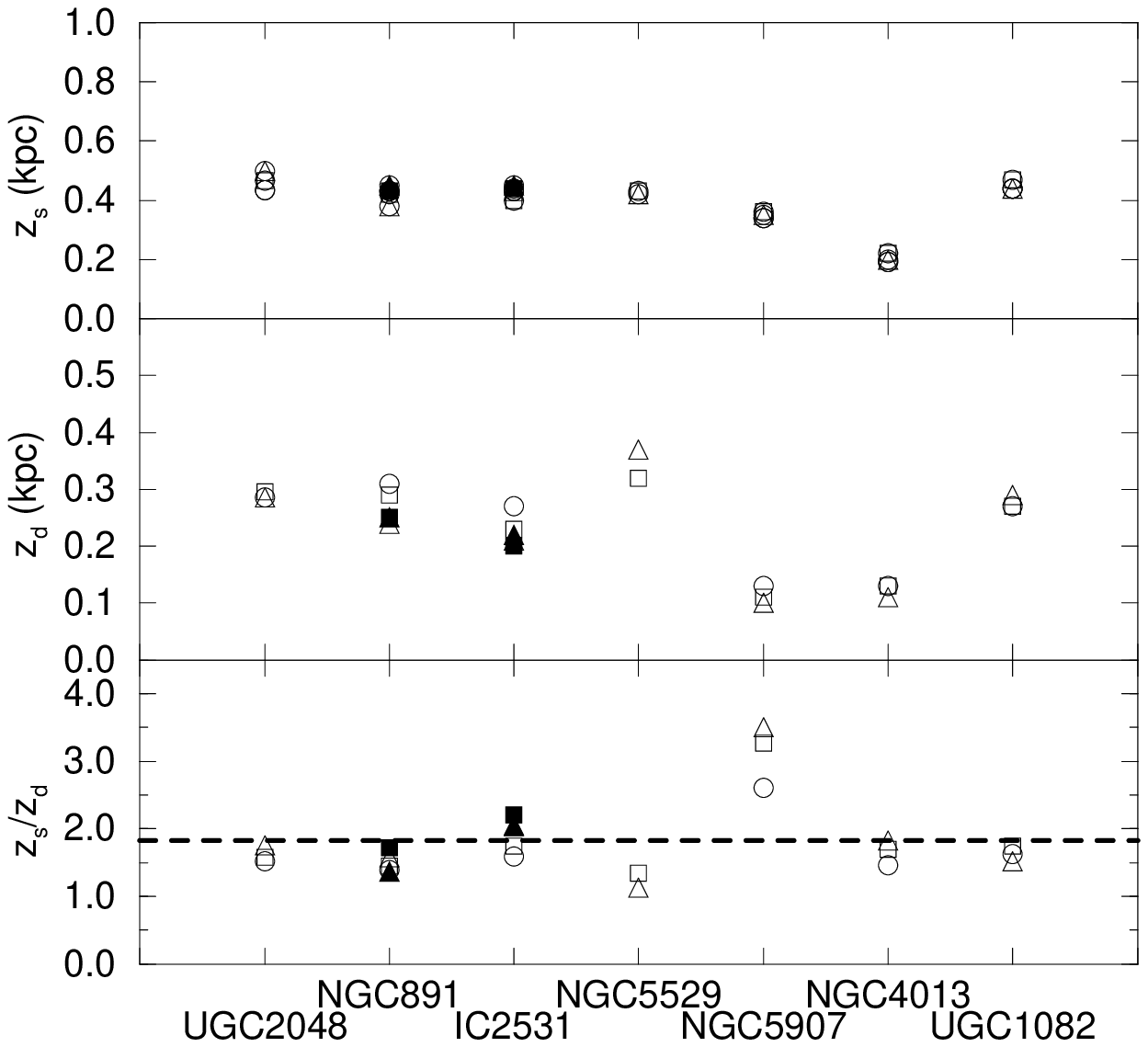}
\caption{The values of $z_s$ (top graph), $z_d$ (middle
graph) and $z_s/z_d$ (bottom graph) for each galaxy.
All lengths are given in kpc.
For the explanation of the symbols see Fig. 8.
The dashed line in the last graph gives the mean value of the ratio
calculated by using the V-band data.}
\end{figure}
\begin{figure}
\epsfxsize=8.9cm
\epsfbox [93 49 483 389]{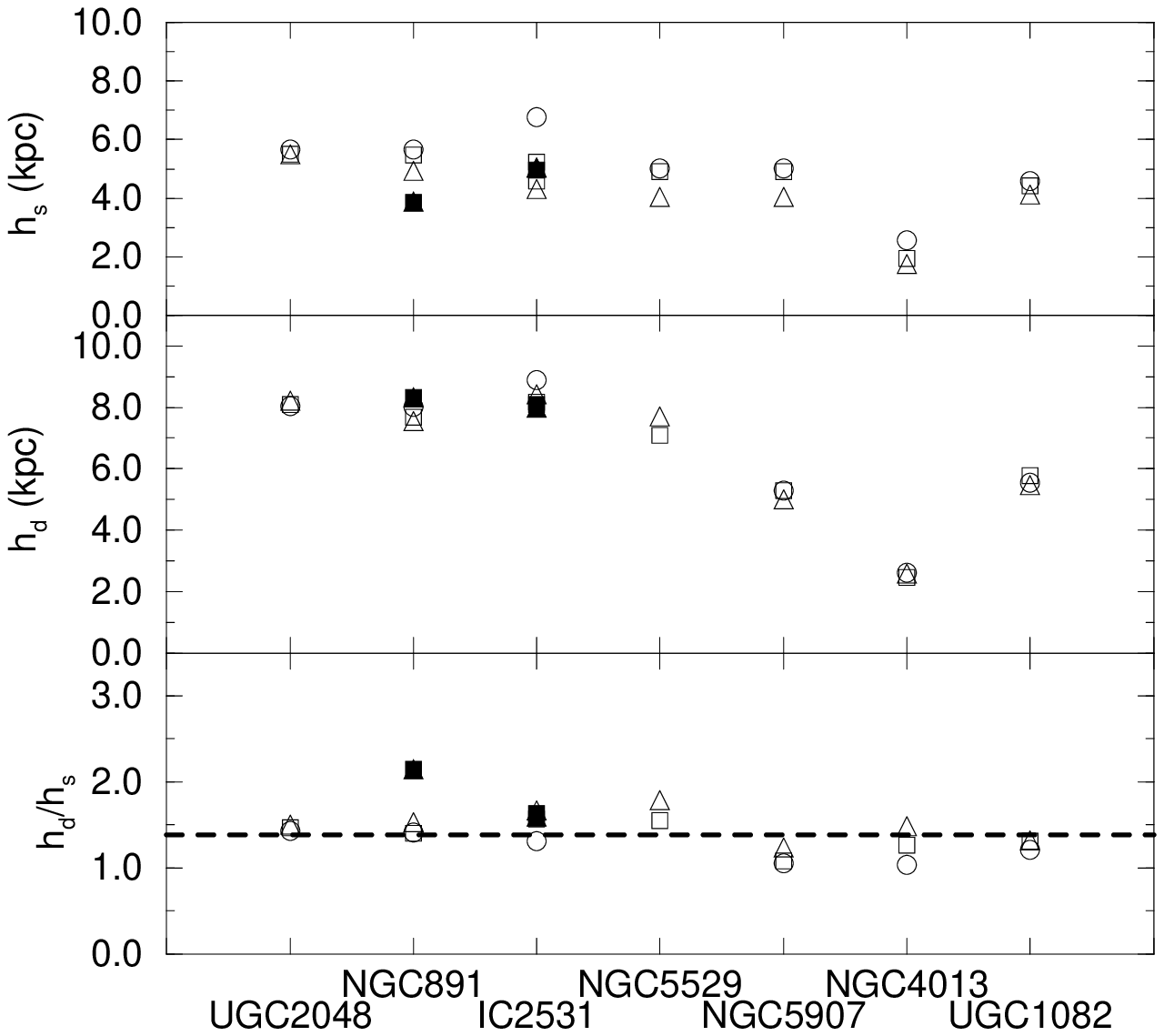}
\caption{
The values of $h_s$ (top graph), $h_d$ (middle
graph) and $h_s/h_d$ (bottom graph) for each galaxy.
All lengths are given in kpc.
For the explanation of the symbols see Fig. 8.
The dashed line in the last graph gives the mean value of the ratio
calculated by using the V-band data.}
\end{figure}
\begin{figure}
\epsfxsize=8.9cm
\epsfbox [93 49 483 389]{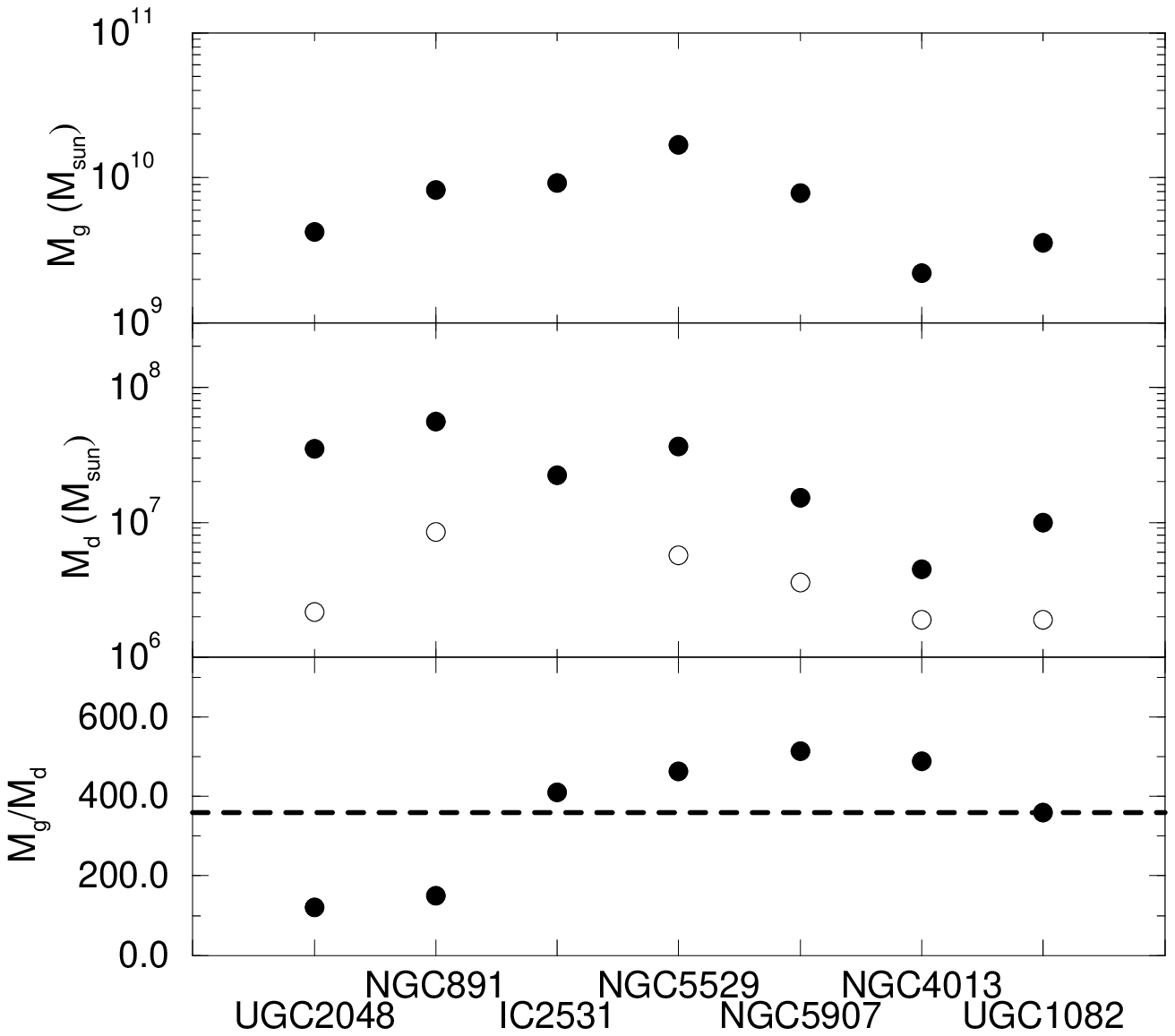}
\caption{
The values of $M_g$ (top graph), $M_d$ (middle
graph) and $M_g/M_d$ (bottom graph) for each galaxy.
All the masses are in units of $M_{\sun}$.
For $M_d$ we give a direct comparison between
the values as calculated by our model (filled circles)
and those calculated using the IRAS fluxes (open circles)
wherever IRAS data were available.
The dashed line in the last graph gives the mean value of the
gas-to-dust mass ratio.}
\end{figure}

\section{Discussion}

Having analyzed a total of seven galaxies, we are now able to draw
some general conclusions concerning the dust and stellar distribution
characteristics of spiral galaxies. 
In order to have a consistent set of parameters
for all the galaxies, we reanalyzed the galaxy NGC 891 (already presented
in Paper II), but now using the $R^{1/4}$ law for the bulge stellar
distribution instead of the modified Hubble profile that was used
in Paper II. The values derived from the modelling are
presented in Table 7 of the present paper.

We have quantified the opacity of the galaxies in terms of their central
face-on optical depth $\tau^f$. {\it For all the galaxies we have found
central face-on 
optical depths with values less than one in all the bands that we analyzed,
indicating that they would appear to be transparent when seen face-on,
despite their obvious dust lane in their almost edge-on orientation
seen in the observations}. The values of the optical depth in all bands are
plotted in Fig. 8 for each galaxy.
Different symbols indicate different bands. In particular,
open circles correspond to the B-band, open squares to the V-band,
open triangles to the I-band, solid squares to the J-band and solid triangles
to the K-band.

It is of interest to see how the scaleheight of the stars 
is related to that of the dust. We show this in Fig. 9,
where the parameters $z_s$ and $z_d$ (in kpc) as well as their ratio
are plotted for each galaxy. The symbols have the same meaning as in Fig. 8.
For the case of UGC 2048 (analyzed in Paper I), we give half 
of the derived scaleheight values since, as mentioned in Paper I, 
it's distance has probably been overestimated by a factor of 2.
Doing the statistics in the V-band data we derive a mean ratio
of $z_s/z_d = 1.8 \pm 0.6$ implying that the dust is more concentrated
to the plane of the disk with respect to the stars.
This was of course, evident from the prominent dust lanes.
This mean value of $z_s/z_d$ is plotted as a horizontal dashed line.
One can also notice that there is not a significant trend of
$z_s$ with wavelength. This means that the main body of the disk
is dominated by the old stellar population, at least at
high $z$ distances, giving roughly the same scaleheight in all
bands.

Of particular interest is the distribution of dust in the
radial direction relative to the stars, as can be seen from the values
that were derived for the scalelengths. 
In order to have a better view of how the scalelengths of the stars 
and the dust are related we give a plot (Fig. 10) similar to that 
of the scaleheights described above. The symbols here, have exactly
the same meaning as that in Fig. 8. As with the scaleheights of
UGC 2048, we adopt half the values of the scalelengths derived from
the model. The statistics that were done in the V-band, give a mean ratio of
$h_d/h_s = 1.4 \pm 0.2$. This value is shown
as a dashed horizontal line.
{\it In all seven galaxies we have found that the dust is
radially more extended than the stars.}
This result
has recently been confirmed
for our Galaxy using COBE/DIRBE data (Davies et al. 1997a).
Analysis of ISO data for several galaxies
(Davies et al. 1997b, Alton et al. 1998a),
as well as SCUBA observations of NGC 891 in 450$\mu m$ and 850$\mu m$ 
(Alton et al. 1998b) also
indicate a more extended radial distribution for the dust 
than for the stars.
Another work that also supports this,
is that of Lequeux \& Guelin (1996) where a systematic reddening of
background galaxies is found in the outer regions of the galaxy of
Andromeda, suggesting the existence of dust even beyond the optical disk.

It is important to notice at this point the deviations with
wavelength shown in Figs. 9 and 10 for the dust parameters.
These deviations, which are mainly due to the different appearance of the
clumpy nature of the galaxies in different bands, give
a better estimate of the real uncertainty which for 
$z_d$ is $\sim 0.05$ kpc and for $h_d$ is $\sim 0.5$ kpc.

An interesting result that comes out of our model analysis is the dust
mass calculation. For the two galaxies analyzed in the previous papers
(UGC 2048; Paper I and NGC 891; Paper II) we find a dust mass
which gives a gas-to-dust mass ratio very close to that derived for our Galaxy
(the values of the dust mass for NGC 891 using the R$^{1/4}$ law profile
becomes $5.6 \times 10^7 M_{\sun}$ while the gas-to-dust mass ratio is
now 150). About the same picture holds for the rest of the galaxies
modelled in this paper. A better picture of this is given in Fig. 11,
where the gas mass (top graph) and the dust mass (middle graph)
are plotted (in units of $M_{\sun}$) for each galaxy. For the dust mass
(middle graph) we give the model calculations
(filled circles) as well as the calculations using the IRAS
data (open circles). The gas-to-dust
mass ratio (using the dust mass derived from the model)
is also given for each galaxy (bottom graph).
A mean value of this ratio is $M_g/M_d = 360 \pm 160$ 
(close to the value of $\sim 160$ adopted for our Galaxy,
see Spitzer 1978; Sodroski et al. 1994) and is plotted as a horizontal
dashed line in the last graph.
{\it For all the galaxies we have found a dust mass about an order of a magnitude
more than that calculated using the IRAS fluxes, which proves the
existence of a cold dust component that has gone undetected by IRAS.}
This is evident from the middle graph of Fig. 11, where the two dust masses
(the one calculated from our model and the other calculated using the
IRAS fluxes) are directly compared.
Our extinction model though, being independent of the dust temperature
(both warm and cold dust contribute to the extinction of light),
is able to calculate the whole amount of dust contained within
the galaxies.
Strong support to this argument (that vast amounts of cold dust
exists in spiral galaxies) is now given by recent far infrared observations
in longer wavelengths, where the cold dust is now seen in emission (Alton et al.
1998a, 1998b; Bianchi et al. 1998; Haas et al. 1998; Kr\"{u}gel et al. 1998).

Having calculated the values of $A_\lambda/A_V$ (see Sect. 4) for the galaxies
modelled in this paper (NGC 891 values are 
$\kappa_B/\kappa_V = 1.23$, $\kappa_I/\kappa_V = 0.61$,
$\kappa_J/\kappa_V = 0.25$ and $\kappa_K/\kappa_V = 0.10$
for the case where the R$^{1/4}$ law is used),  
we are able to show
how the values of the relative extinction in each band is compared with
the values derived for our Galaxy (Rieke \& Lebofsky 1985). This is
done in Fig. 12, where the values of the extinction ratios
$A_\lambda/A_V$ are plotted as a function of the inverse wavelength
($1/\lambda$). Each symbol here represents a different galaxy
as shown inside the graph.
One can see that {\it the extinction law derived for all the galaxies
is in very good agreement with that measured for our Galaxy, indicating
a universal dust behaviour}.

\begin{figure}[t]
\epsfysize=7.3cm
\epsfbox [99 64 488 385]{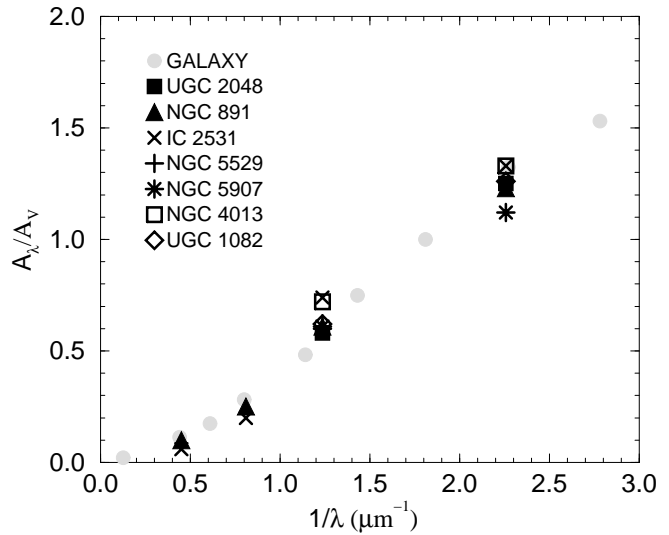}
\caption{The observed (gray circles) values of $A_{\lambda}/ A_V$ for our
Galaxy and the values calculated for all the galaxies analyzed.
The symbols are explained in the plot.}
\end{figure}

\begin{figure}
\epsfysize=11.5cm
\epsfbox [0 0 380 506]{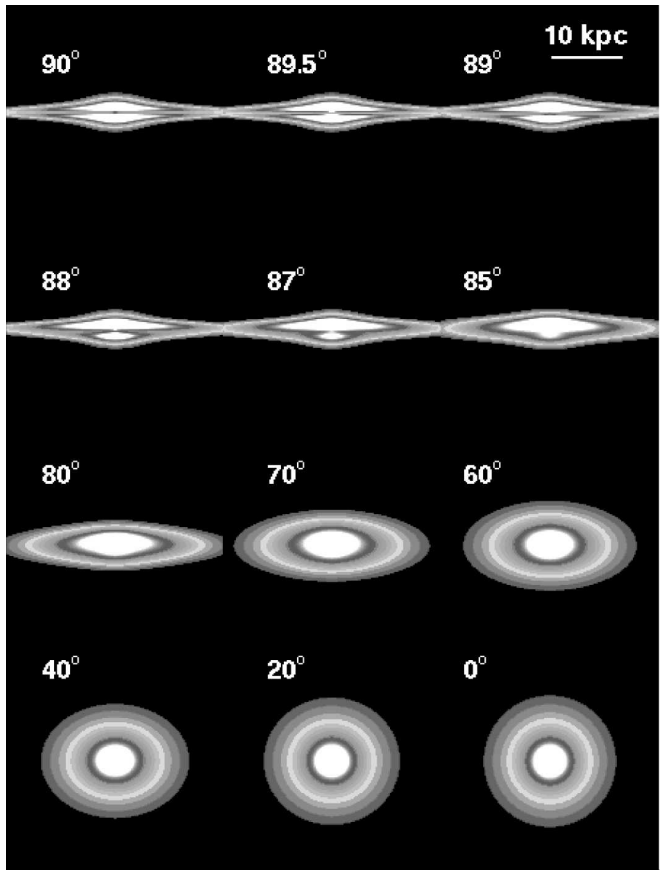}
\caption{A typical model galaxy in the B-band shown at various
inclination angles (which are indicated on top of each model).
The fainter surface brightness level at the outer part of the galaxy is
25 mag/arcsec$^2$.}
\end{figure}
\begin{figure}[!t]
\epsfysize=11.5cm
\epsfbox [2 2 380 506]{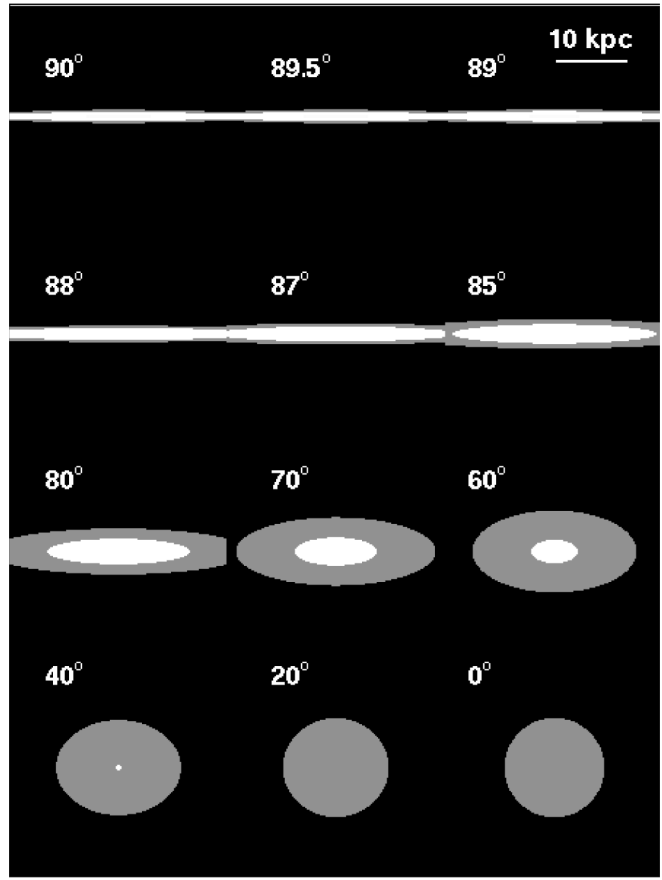}
\caption{Mapping of the optical depth as a function of the
inclination angle. White regions show the optically
thick parts of the galaxy (with $\tau > 1$) while grey
regions show the optically thin parts of the galaxy.}
\end{figure}

\section{Summary}
In this paper we have done the modelling of the dust and stellar
content of five spiral galaxies (NGC 4013, IC 2531, UGC 1082,
NGC 5529 and NGC 5907). With the two galaxies that have already
been analyzed in previous work (UGC 2048 and NGC 891),
we are able to draw some general conclusions and thus obtain a 
good picture of how a typical late-type spiral galaxy
looks like. Assuming that the dust and the stellar disk
follow an exponential 3D distribution and that the bulge
is described by an $R^{1/4}$ law profile, the
typical parameters (derived from the mean values)
that describe the galaxy in the B-band are
$z_s \approx 0.4$ kpc,  $h_s \approx 5.0$ kpc and $I_s \approx 20$ mag/arcsec$^2$ for 
the stellar disk and
$z_d \approx 0.5 z_s$ and $h_d \approx 1.4 h_s$ for the dust disk.
A mean B-band central face-on optical depth is $\tau^f \approx 0.8$.
For the bulge, being more dependent on the morphological type
of the galaxy, we consider the values $b/a = 0.5$ for the ellipticity,
$R_e = 1.5$ kpc for the effective radius and $I_b \approx 12$ mag/arcsec$^2$ for
the central edge-on surface brightness.
Such a galaxy is shown in Fig. 13 in various inclination angles,
ranging from exactly edge-on (90 $\degr$) to face-on (0 $\degr$).
The faintest surface brightness level is 25 mag/arcsec$^2$.
One thing that can clearly be seen is the existence of a dust
lane down to almost $85 \degr$. The decrease in the optical diameter
(going from edge-on to face-on) is also very obvious.
This is the effect of integrating along larger
paths of light in the edge-on configuration than in the face-on one.
In Fig. 14 we show how the ``imagei'' of the optical depth of such a galaxy
varies with inclination angle. In each model-galaxy shown in this
figure, the white regions show the parts of the galaxy that are 
optically thick (with $\tau > 1$) and grey regions show the
optically thin parts of the galaxy.
It is evident that the galaxy is optically thick, at least in the central
regions, down to almost inclination angle of $60 \degr$, while the
galaxy becomes totally transparent when viewed face-on.

\begin{acknowledgements}
We are greatful to P. Alton for calculating the dust masses of the galaxies
using the IRAS fluxes.
This paper has been benifit a lot by the comments of R. Corradi.
We also thank I. Papadakis, G. Paterakis,
F. Mavromatakis and A. Misiriotis for stimulating
discussions.
This research has been supported in part by a Greek-British Joint Research
Program and by a P.EN.E.D. Program of the General Secretariat of Research
and Technology of Greece.
YIB was supported by Creative Research Initiatives Program of
the Korean Ministry of Science and Technology and also by Yonsei
University Research Grant.
Skinakas Observatory is a collaborative project of the University of
Crete, the Foundation for Research and Technology-Hellas and
the Max-Planck-Institut f\"ur Extraterrestrische Physik.
\end{acknowledgements}


\begin{thebibliography}{}
\bibitem[]{}
Alton P.B., Trewhella M., Davies J.I., 1998a, A\&A, 335, 807
\bibitem[]{}
Alton P.B., Bianchi S., Rand R.J., Xilouris E.M., Davies J.I., 
Trewhella M., 1998b, ApJ, 507, L125
\bibitem[]{}
Barnaby D., Thronson H.A., Estep G.M., 1993, BAAS, 183, \#79.02
\bibitem[]{}
Barnaby D., Thronson H.A. 1992, AJ, 103, 41
\bibitem[]{}
Bianchi S., Alton P.B., Davies J.I., Trewhella M., 1998, MNRAS,
298, 49
\bibitem[]{}
Bottema R., 1995, A\&A, 295, 605
\bibitem[]{}
Bottema R., 1996, A\&A, 306, 345
\bibitem[]{}
Bruzual G.A., Magris G.C., Calvet N., 1988, ApJ, 333, 673
\bibitem[]{}
Davies J.I., Trewhella M., Jones H., et al., 1997a, MNRAS, 288, 679
\bibitem[]{}
Davies J.I., Alton P. B., Trewhella M., Bianchi S., 1997b,
MNRAS, submitted
\bibitem[]{}
Devereux N.A., Young J.S., 1990, ApJ, 359, 42
\bibitem[]{}
Disney M.J., Davies J.I., Phillipps S., 1989, MNRAS, 239, 939
\bibitem[]{}
Domingue D.L., Keel W.C., White R.E., 1998, ApJ, submitted
\bibitem[]{}
Dumke M., Braine J., Krause M., et al., 
1997, A\&A, 325, 124
\bibitem[]{}
Florido E., Prieto M., Mediavilla E., Sanchez-Saavedra M.L.,
1991, A\&A, 242, 301
\bibitem[]{}
Giovanelli R., Haynes M. P., 1993, AJ, 105, 1271
\bibitem[]{}
Giovanelli R., Haynes M. P., Salzer J.J., et al., 1994, AJ, 107, 2036
\bibitem[]{}
Giovanelli R., Haynes M. P., Salzer J.J., et al., 1995, AJ, 110, 1059
\bibitem[]{}
Gomez de Castro A.I., Garcia-Burillo S., 1997, A\&A, 322, 381
\bibitem[]{}
Gonz\'{a}lez R.A., Allen R.J., Dirsch B., at al., 1998
ApJ, 506, 152
\bibitem[]{}
Graham J. A., 1982, PASP, 94, 244
\bibitem[]{}
Haas M., Lemke D., Stickel M., et al., 1998, A\&A, 338, L33
\bibitem[]{}
Henyey L.G., Greenstein J.L., 1941, ApJ, 93, 70
\bibitem[]{}
Huchtmeier W.K., Richter O.G., 1989, A general catalog of $H_I$
observations of galaxies, New York, Springer-Verlag
\bibitem[]{}
Just A., Fuchs B., Wielen R., 1996, A\&A, 309, 715
\bibitem[]{}
Kodaira K., Yamashita T., 1996, PASJ, 48, 581
\bibitem[]{}
Kr\"{u}gel E., Siebenmorgen R., Zota V., Chini R., 1998,
A\&A, 331, L9
\bibitem[]{}
Kuchinski L.E., Terndrup D.M., Gordon K.D., Witt A.N., 1998,
AJ, 115, 1438
\bibitem[]{}
Kylafis N.D., Bahcall J.N., 1987, ApJ, 317, 637 (KB87)
\bibitem[]{}
Lequeux J., Guelin M., 1996, New extragalactic perspectives
in the New South Africa, Kluwer Academic Publishers
\bibitem[]{}
Misiriotis A., Kylafis N., Papamastorakis J., Xilouris E.,
1999, in preparation
\bibitem[]{}
Moriondo G., Giovanelli R., Haynes M.P., 1998, A\&A, 338, 795
\bibitem[]{}
Morrison H.L., Boroson T.A., Harding P., 1994, AJ, 108, 1191
\bibitem[]{}
Moshir M., Conrow G., McCallon H., et al., 1990, Infrared
astronomical satellite catalogs, the faint source catalog.
\bibitem[]{}
Rieke G.H., Lebofsky M.J., 1985, ApJ, 288, 618
\bibitem[]{}
R\"{o}nnback J., Shaver P.A., 1997, A\&A, 322, 38
\bibitem[]{}
Shaw M., Dettmar R.-J., Barteldrees A., 1990, A\&A, 240, 36
\bibitem[]{}
Sodroski T.J., Bennett C., Boggess N., et al., 1994, ApJ,
428, 63
\bibitem[]{}
Spitzer L., 1978, Physical Processes in the Interstellar Medium,
New York, Wiley-Interscience
\bibitem[]{}
van der Kruit P.C., Searle L., 1982, A\&A, 110, 61
\bibitem[]{}
Wainscoat R.J., Freeman K.C., Hyland A.R., 1989, A\&A, 337, 163
\bibitem[]{}
Xilouris E.M., Kylafis N.D., Papamastorakis J., Paleologou E.V.,
Haerendel G., 1997, A\&A, 325, 135 (Paper I)
\bibitem[]{}
Xilouris E.M., Alton P.B., Davies J.I., et al., 1998, A\&A, 331, 894 (Paper II)
\bibitem[]{}
Zombeck M.V., 1990, Handbook of space astronomy and astrophysics,
Cambridge University Press
\end{thebibliography}
\end{document}